\documentclass[a4paper]{scrartcl}

\usepackage{graphicx}
\usepackage{subfig}
\usepackage{amsmath}
\usepackage{amssymb}
\usepackage{nicefrac}
\usepackage[T1]{fontenc}
\usepackage[utf8]{inputenc}
\usepackage[affil-it]{authblk}
\usepackage[colorlinks=true, linkcolor=red, citecolor=blue]{hyperref}
\usepackage[sort&compress,numbers]{natbib}

\begin{document}

\date{}
\author[1]{Asis Pattisahusiwa}
\author[2]{Acep Purqon}
\author[3]{Sparisoma Viridi}
\affil[1]{Master Program in Physics, Institut Teknologi Bandung, Jl. Ganesha 10, Bandung 40132, Indonesia}
\affil[2]{Geophysics and Complex System Research Division, Department of Physics, Institut Teknologi Bandung, Jl. Ganesha 10, Bandung 40132, Indonesia}
\affil[3]{Nuclear physics and Biophysics Research Division, Department of Physics, Institut Teknologi Bandung, Jl. Ganesha 10, Bandung 40132, Indonesia}

\title{\Large\bfseries Hydrostatic Simulation of Earth's Atmospheric Gas Using Multi-particle Collision Dynamics\footnote{%
Presented at Padjadjaran Earth Dialogue: International Symposium on Geophysical Issues (PEDISGI), 8--10 June 2015. Submitted to IOP Conference Series: Earth and Environmental Science.}}
\maketitle
	
	\begin{abstract}
		Multi-particle collision dynamics (MPCD) is a mesoscopic simulation method to simulate fluid
		particle-like flows. 
		MPCD has been widely used to simulate various problems in condensed matter. 
		In this study, hydrostatic behavior of gas in the Earth's atmospheric layer is simulated by using MPCD method. 
		The simulation is carried out by assuming the system under ideal state and is affected only by gravitational force.
		Gas particles are homogeneous and placed in 2D box.
		Interaction of the particles with the box is applied through implementation of boundary conditions (BC).
		Periodic BC is applied on the left and the right side, specular reflection on the top side, while bounce-back on the bottom side. 
		Simulation program is executed in Arch Linux and running in notebook with processor Intel i5 @2700 MHz with 10 GB DDR3 RAM. 
		The results show behaviors of the particles obey kinetic theory for ideal gas when gravitational acceleration value is proportional to the particle mass. 
		Density distribution as a function of altitude also meets atmosphere's hydrostatic theory.
	\end{abstract}
	
	\section{Introduction}		

		Atmosphere is a layer that surrounds the earth and is mostly composed of molecular gases and plasma in upper atmosphere.
		Atmosphere is interesting to study since it has a lot of complex phenomena and still under active study.
		The complexity is mostly contributed by the interaction of fluids, such as the interaction between ocean and neutral gas, or between neutral gas and plasma.
		Hence, any computational fluid dynamics methods can be used to observe this complexity.

		In the other hand, Multi-particle collision dynamics (MPCD) that was first proposed by Malevanets and Kapral in 1999~\citep{malevanets1999mesoscopic} offers a simple computational method to model fluids dynamics.
		In MPCD method, as well as in the other particle-based methods, the fluids are assumed as a collection of particle-like objects with zero volume.
		MPCD algorithm is similar to Direct Simulation Monte Carlo (DSMC)~\citep{bird1994molecular}, the difference is only on the collision mechanism among the particles.
		The collision in DSMC method is represented by binary-collision between the pair of particles;
		while in MPCD, the collision is represented by stochastic rotation dynamics.
		MPCD with this collision mechanism is known as Stochastic Rotation Dynamics (SRD).
		MPCD with Anderson Thermostat (MPCD-AT) can be used as an alternative to SRD mechanism~\citep{gompper2009multi}.
		The method has been widely used to simulate various problems in the complex fluids.
		For example, simulation of solute-solvent dynamics~\citep{malevanets2000solute,allahyarov2002mesoscopic}, claylike colloids~\citep{hecht2005simulation}, star polymers in solution~\citep{singh2014hydrodynamic}, and flow-induced polymer translocation~\citep{nikoubashman2010flow}.

		As far as we know, MPCD method has never been used in the atmospheric hydrodynamics or hydrostatics problems.
		So, the method will be tested for its ability when applied to atmospheric problems.
		For this purpose, MPCD will be applied to a simple problem, simulation of behavior of atmospheric neutral gas in hydrostatic equilibrium.
		The standard of hydrostatic atmospheric model used in this study will be described in the \autoref{sec:atmosphere}, while details of the method and simulation mechanism will be explained in \autoref{sec:mpcd}.

	\section{Standard Atmospheric Model}
	\label{sec:atmosphere}
		Based on U.S. Standard Atmosphere 1976 \citep[pp.~6--7]{coesa1976standard}, the atmosphere up to altitude $86$ km is assumed to be dry, homogeneously mixed, and obeys ideal gas law.
		The effect of homogeneity leads to composition of the atmosphere up to this altitude have uniform molecular weight, $M_0=$ $28.96443$ $\text{kg kmol}^{-1}$.
		The atmospheric composition is dominated by nitrogen gas $N_2$ with $78.084$ $\%$, oxygen gas ($20.9476$ $\%$), and traces of $Ar$, $CO_2$, $Ne$, $He$, $Kr$, $Xe$, $CH_4$, and $H_2$.

		Atmospheric hydrostatic state can be described by using ideal gas law.
		Supposed atmospheric temperature $T$, particle density $\rho$, and molecular weight $M_0$, the atmospheric pressure is given by
		\begin{align}
			P = \dfrac{\rho R^\ast T}{M_0}.
			\label{eq:tekanan}
		\end{align}
		Profile of particle density for any altitude can be deduced from equation~\eqref{eq:tekanan} and is given by
		\begin{align}
			\rho(z) = \rho_0 \exp\left(-\dfrac{(z - z_0)}{H}\right),
			\label{eq:rho.hidrostatik}
		\end{align}
		where $z_0$, $z$ are initial and current altitudes, respectively.
		Scale height $H$ is defined by
		\begin{align*}
			H = \dfrac{R^\ast T}{g},
			\label{eq:model.scale.height}
		\end{align*}
		where $R^\ast$, $g$ are universal gas constant and gravitational acceleration respectively.
		If we assume temperature $T = T_0=$ $288.15$ K, then the scale height is $8.4334$ km.

	\section{Multi-particle Collision Dynamics}
	\label{sec:mpcd}
		
		Simulation step in MPCD algorithm is divided into two steps, streaming and collision.
		In the streaming step, particle position is updated based on its velocity calculated in the collision step.
		The particle position for any time step is calculated by using
		\begin{align*}
			\vec{r}_i(t+\delta t) = \vec{r}_i(t) + \vec{v}_i(t)\delta t + \dfrac{1}{2}\vec{g}\delta t^2.
		\end{align*}
		In this equation, $i$ express the index of the particles, $\vec{r}(t+\delta t)$ is position at time $t + \delta t$, while $\vec{r}(t)$, $\vec{v}(t)$ are position and velocity at time $t$ respectively.

		In the collision step, particles are sorted into a collision cell with lateral size $a$.
		MPDC algorithm allows the particles to interact with each other in the same cell only.
		The interaction itself is represented by stochastic rotation matrix with random angle for every time step.
		Particle velocity after interaction is calculated by using
		\begin{align*}
			\vec{v}_i(t + \delta t) = \vec{u} + \mathbf{R}\delta\vec{v}, \qquad \delta\vec{v} = \vec{u} - \vec{v}_i(t),
		\end{align*}
		where $\vec{v}$, $\vec{u}$, $\mathbf{R}$, $\delta\vec{v}$ are velocity after interaction, velocity of center of mass, rotation matrix, and relative velocity between particle and center of mass, respectively.
		The velocity of center of mass is expressed by
		\begin{align*}
			\vec{u} = \dfrac{1}{N_c}\sum\limits_{i}^{N_c}\vec{v}_i(t),
		\end{align*}
		where $N_c$ is number of particle per cell.
		Two-dimensional rotation matrix is calculated by using
		\begin{align*}
			\mathbf{R} = 
			\left[
				\begin{array}{ll}
					\cos(n\theta) & -\sin(n\theta)
					\\
					\sin(n\theta) & \cos(n\theta)
				\end{array}
			\right],
		\end{align*}
		where $n$ is chosen between $+1$ and $-1$ randomly, while $\theta$ in this study is set to $90^\circ$.

		Interaction of the particles with the box is applied through the implementation of boundary
		conditions (BC).
		Periodic BC is applied on the left and the right of the box, while specular reflection and bounce-back is applied on the top and the bottom side respectively.
		Implementation of the last two BC is implemented by modifying tangential and normal components of particle velocity based on
		\begin{align*}
			\vec{v}_t &= \left(2\Gamma - 1\right)\vec{v}_t,
			\\
			\vec{v}_n &= -\vec{v}_n,
			\\
			\vec{v} &= \vec{v}_t + \vec{v}_n,
		\end{align*}
		where $\Gamma=1$ for specular reflection and $\Gamma=0$ for bounce-back.
		
		In the equation \eqref{eq:rho.hidrostatik}, we can see that the density profile is dependent only on altitude.
		Hence, motion of the particles in two-dimensional coordinate is sufficient to achieve this profile.
		In this study, The simulation is performed by assuming that there are $N$ particles with mass $m=1$, position $\vec{r}$ and velocity $\vec{v}$.
		The particles are distributed randomly in two-dimensional box with size $10a\times 20a$, $a=1$.
		The number of collision cells in the box is $200$ and average number of particles per cell is fixed to $5$.
		We set $R^\ast T = 1$, this is led to mean particle speed $v = 1.5958$, and speed of sound $c_s = 1.1832$.
		
		Each particle is given a random initial velocity which follows normal distribution.
		The distribution is generated from Box-M\"uller transformation and expressed by~\citep{lee2006hardware}
		\begin{align*}
			y_1 &= \sqrt{-2\log(u_1)}\cos(2\pi u_2),
			\\
			y_2 &= \sqrt{-2\log(u_1)}\sin(2\pi u_2),
		\end{align*}
		where $u_1$, $u_2$ are uniformly distributed random numbers, and $y_1$, $y_2$ are a pair of normal distributed random numbers.
		The pair of values are used interchangeably to produce the initial velocity for each particle through
		\begin{align*}
			\vec{v}_i = y\sigma + \mu; \qquad y = 
			\left\{%
				\begin{array}{ll}
					y_1, & \text{$i$ even}
					\\
					y_2, & \text{$i$ odd}
				\end{array}
			\right.,%
		\end{align*}
		where $\sigma$ and $\mu$ are standard deviation and mean of the distribution.
		In this study, $\mu=0$, while $\sigma$ is the thermal speed of particle.
		
		Mean free path is set to $0.25a$ and corresponding to collision interval
		\begin{align*}
			\delta t = \lambda t_0; \qquad t_0 = a\sqrt{\dfrac{M_0}{R^\ast T}},
		\end{align*}
		where $t_0$ is unit time.
		Mean free path is taken smaller than lateral size ($\lambda \ll a$) indicates that there are no molecular chaos and Galilean invariant \citep{ihle2001stochastic}.
		This problem can be solved by using grid shifting method \citep{ihle2001stochastic}.
		The method is implemented by multiplying the grid position
		before collision with any random number $b$ and then moving back after collision.
		The random number is chosen in the range $\nicefrac{+a}{2}$ and $\nicefrac{-a}{2}$.
		
		Since the system depends on value of gravitation acceleration, it will be varied in performing the simulation.
		The value is chosen among $1$, $1.5$, $2$, $2.5$, $3$, and $3.5$.
		The simulation itself is carried out until 5000 time steps.
		The density profile is generated from average density value from 2500 to 5000 time steps, while the particle speed distribution is taken from the last time step.

	\section{Results and Discussion}
		
\begin{figure}[!ht]
			\vskip -1em
			\centering
			\subfloat[]
			{
				\includegraphics[width=0.35\textwidth]{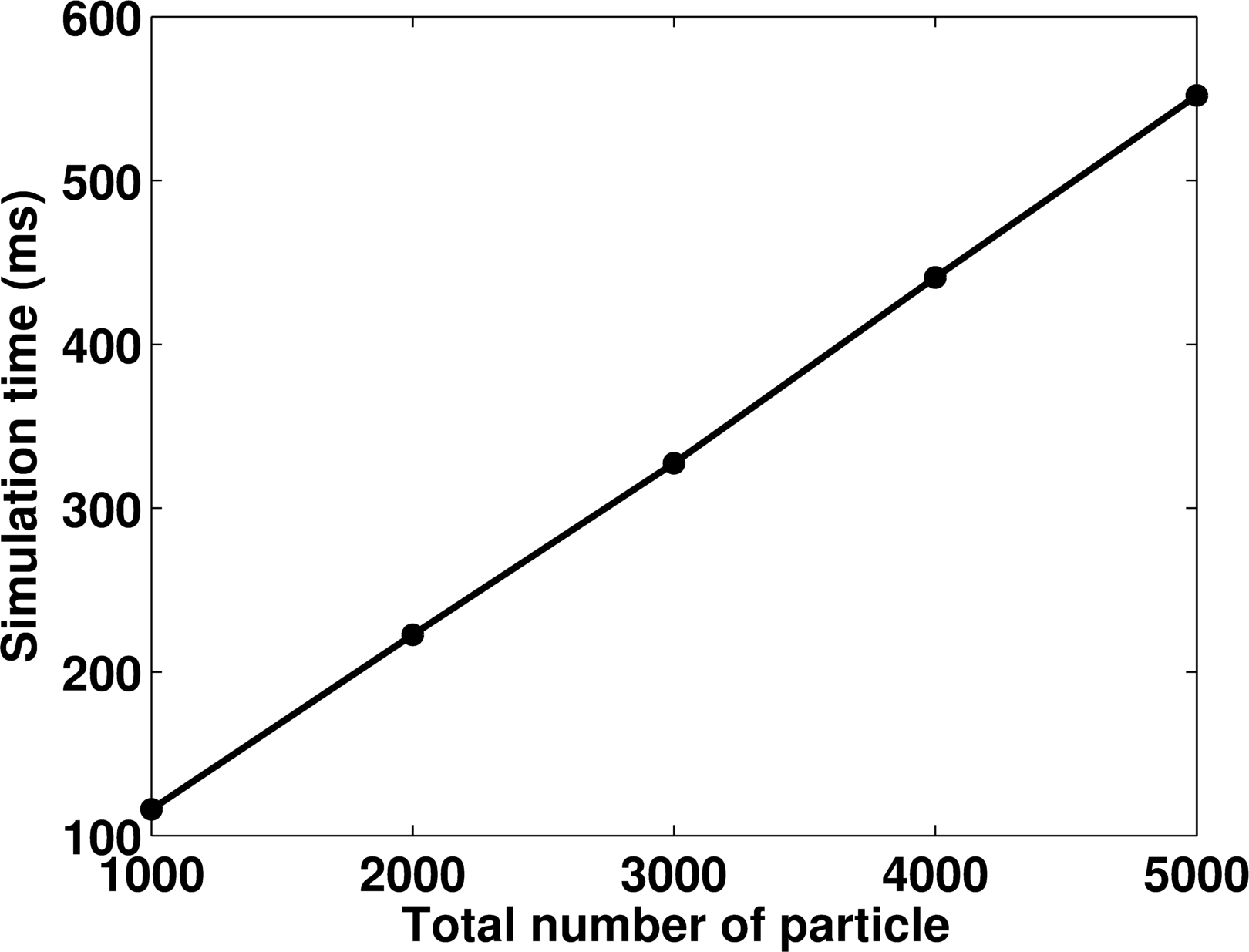}
				\label{img:simulation.time}
			}
			\hfil
			\subfloat[]
			{
				\includegraphics[width=0.35\textwidth]{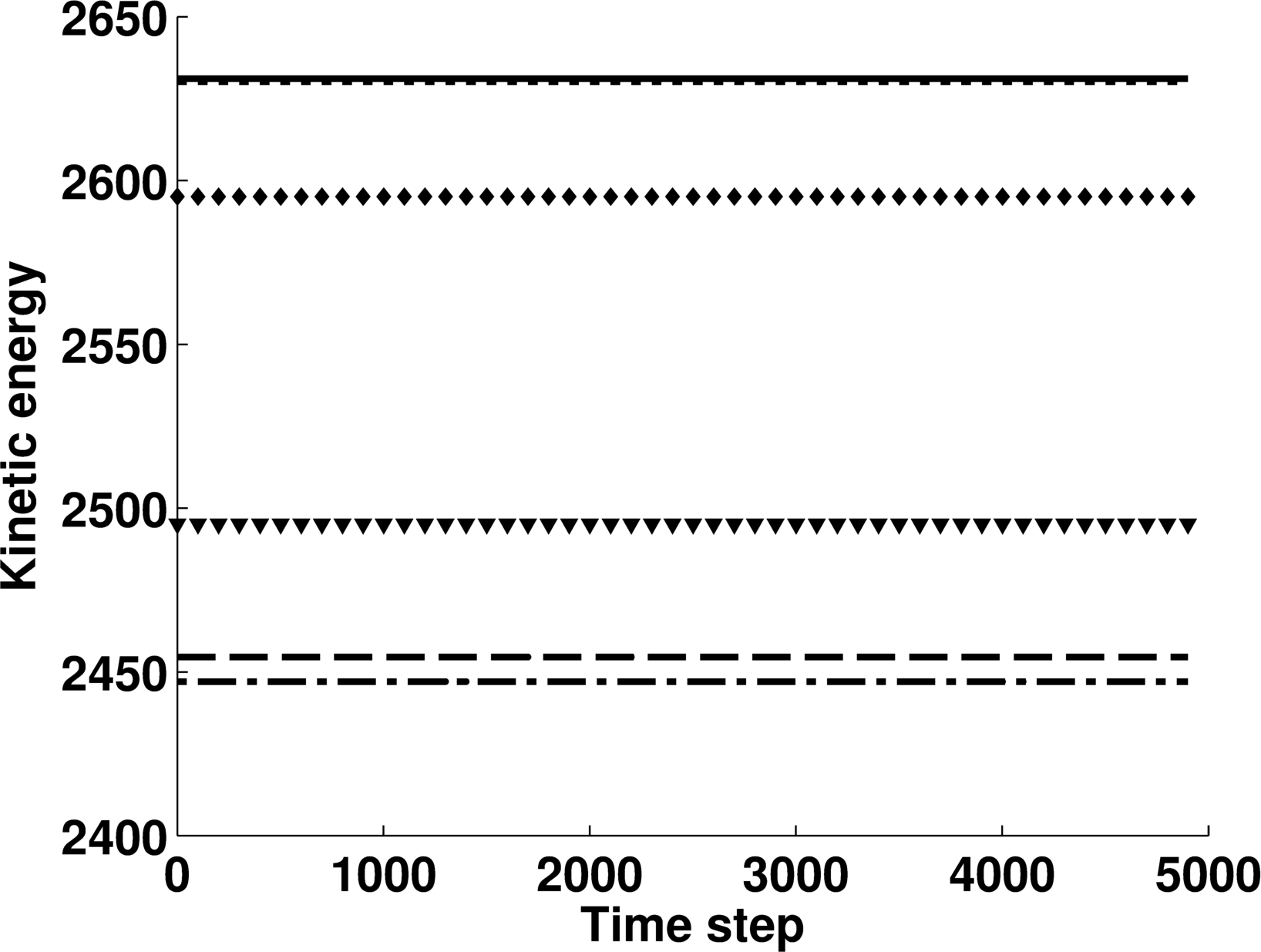}
				\label{img:kinetic}
			}
			\caption{(a) Average simulation time for each total number of particles; (b) Profile of kinetic energy during simulation: $g=1$ (dashed), $g=1.5$ (solid), $g=2$ (dotted), $g=2.5$ (dash-dotted), $g=3$ ($\blacktriangledown$), and $g=3.5$ ($\blacklozenge$).}
		\end{figure}

		Simulation program is built by using Xojo (\url{http://www.xojo.com}) which based on BASIC language.
		The application is programmed and executed in Arch Linux with processor Intel i5 $@2700$ MHz and 10 GB DDR3 RAM.
		Figure \ref{img:simulation.time} shows the average of simulation time for several total number of particles during both streaming and collision steps for each time step.
		Although, We can not compare this result with simulation time of others methods, MPCD is reliable as an alternative of lightweight method for fluid simulation.
		On the figure \ref{img:kinetic}, We can see that the algorithm satisfies energy conservation and therefore also fulfill linear momentum conservation.
	
\begin{figure}[!ht]
			\vskip -1em
			\centering
			\subfloat[$g=1$]
			{
				\includegraphics[width=0.3\textwidth]{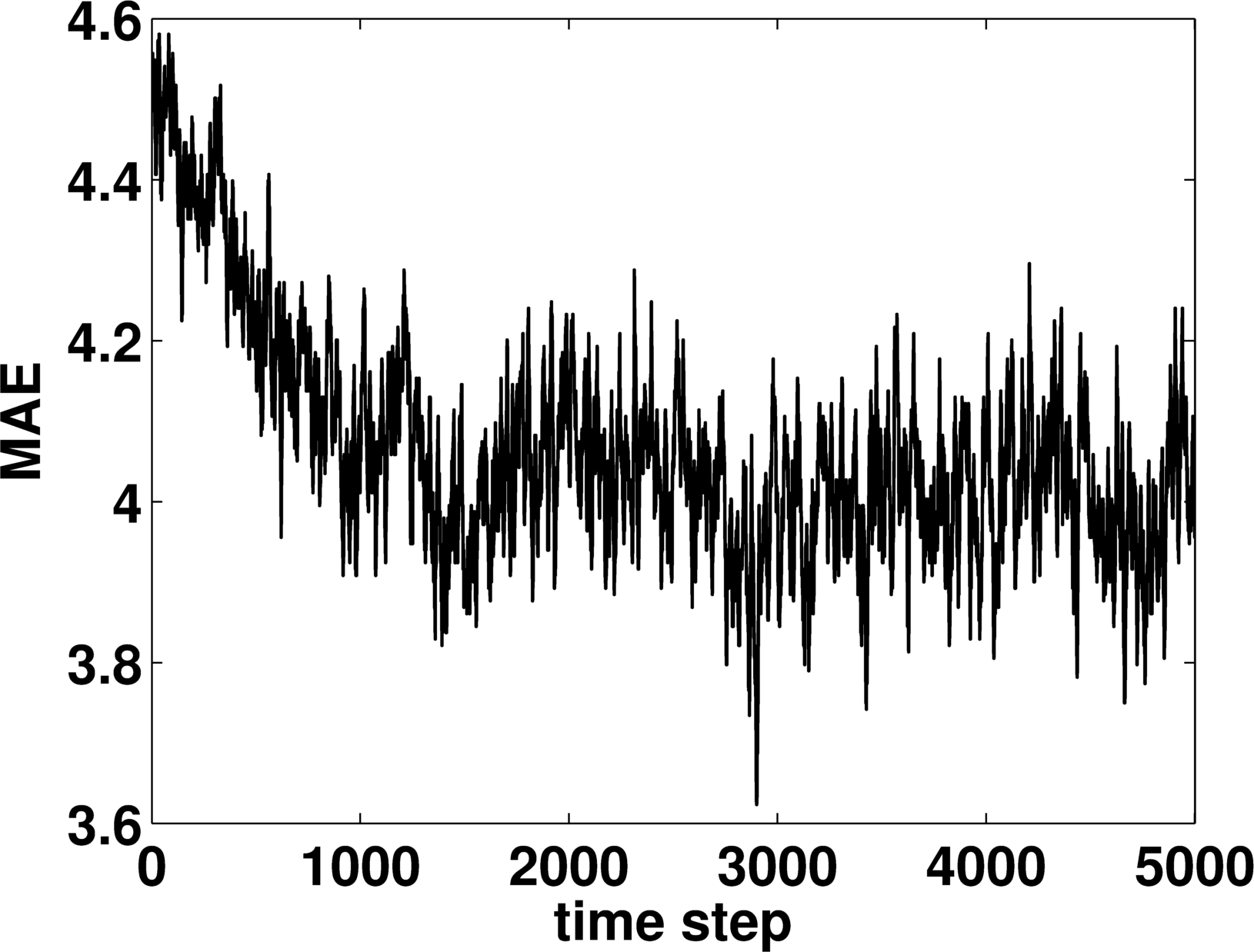}
			}
			\hfil
			\subfloat[$g=1.5$]
			{
				\includegraphics[width=0.3\textwidth]{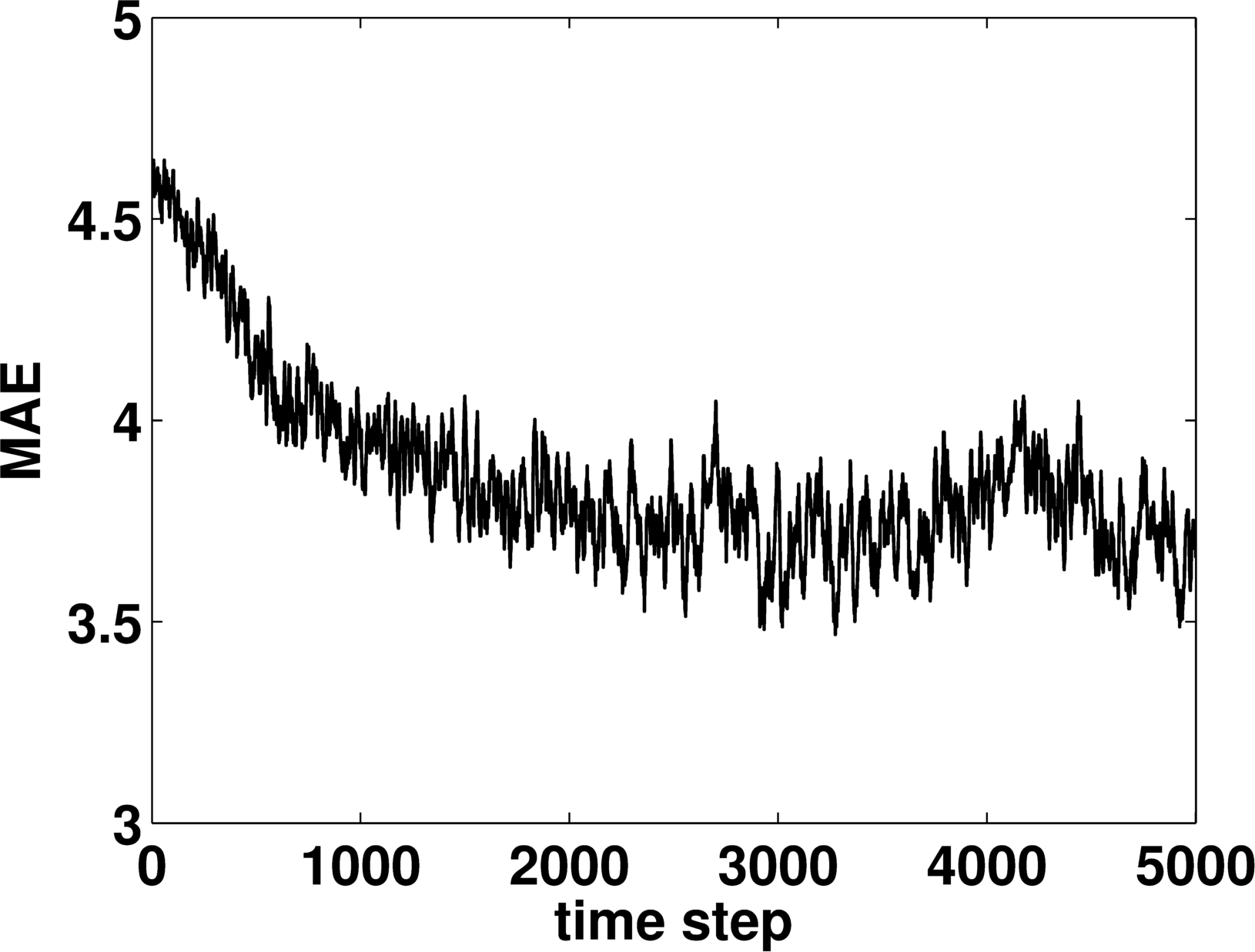}
			}
			\hfil
			\subfloat[$g=2$]
			{
				\includegraphics[width=0.3\textwidth]{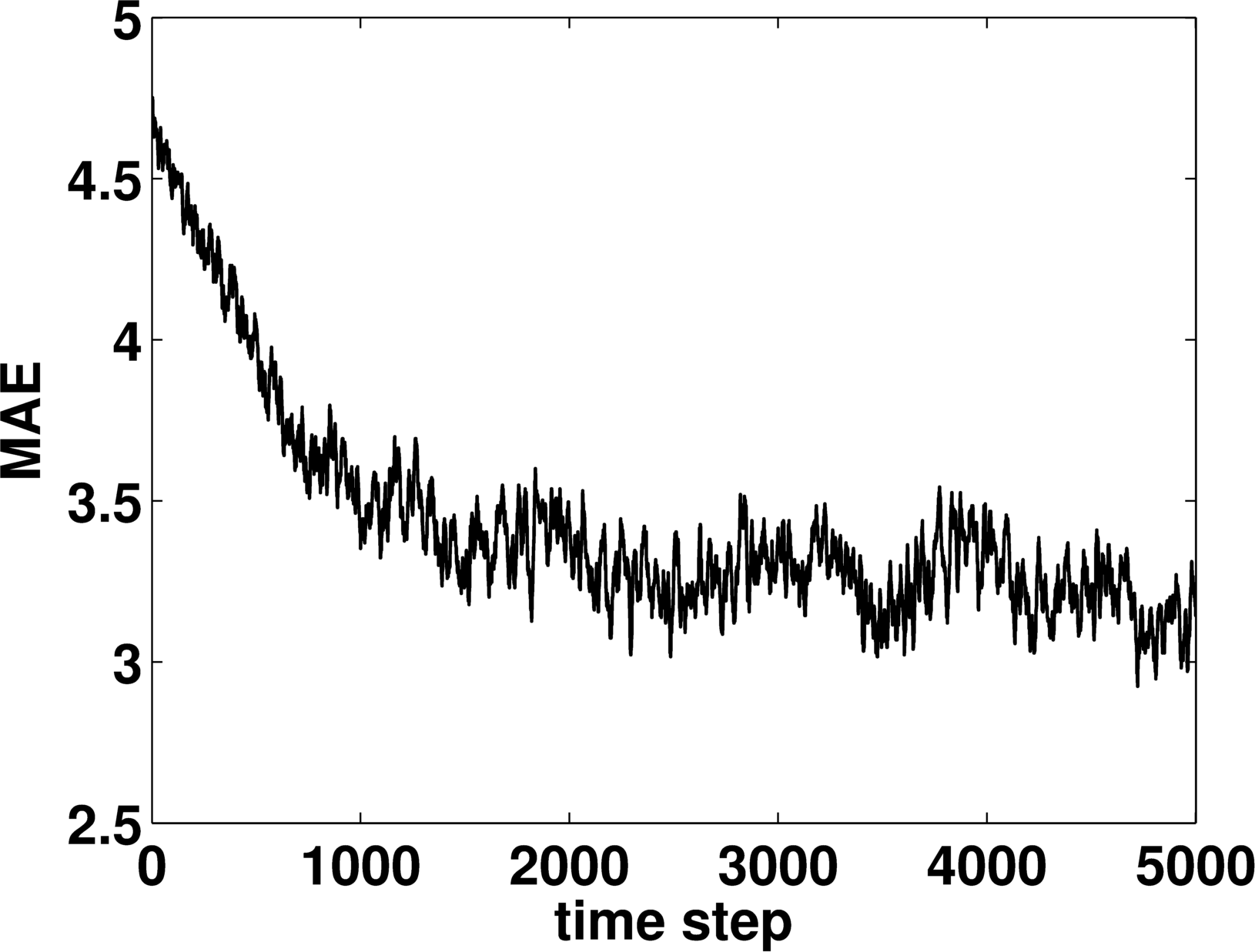}
			}
			\\
			\subfloat[$g=2.5$]
			{
				\includegraphics[width=0.3\textwidth]{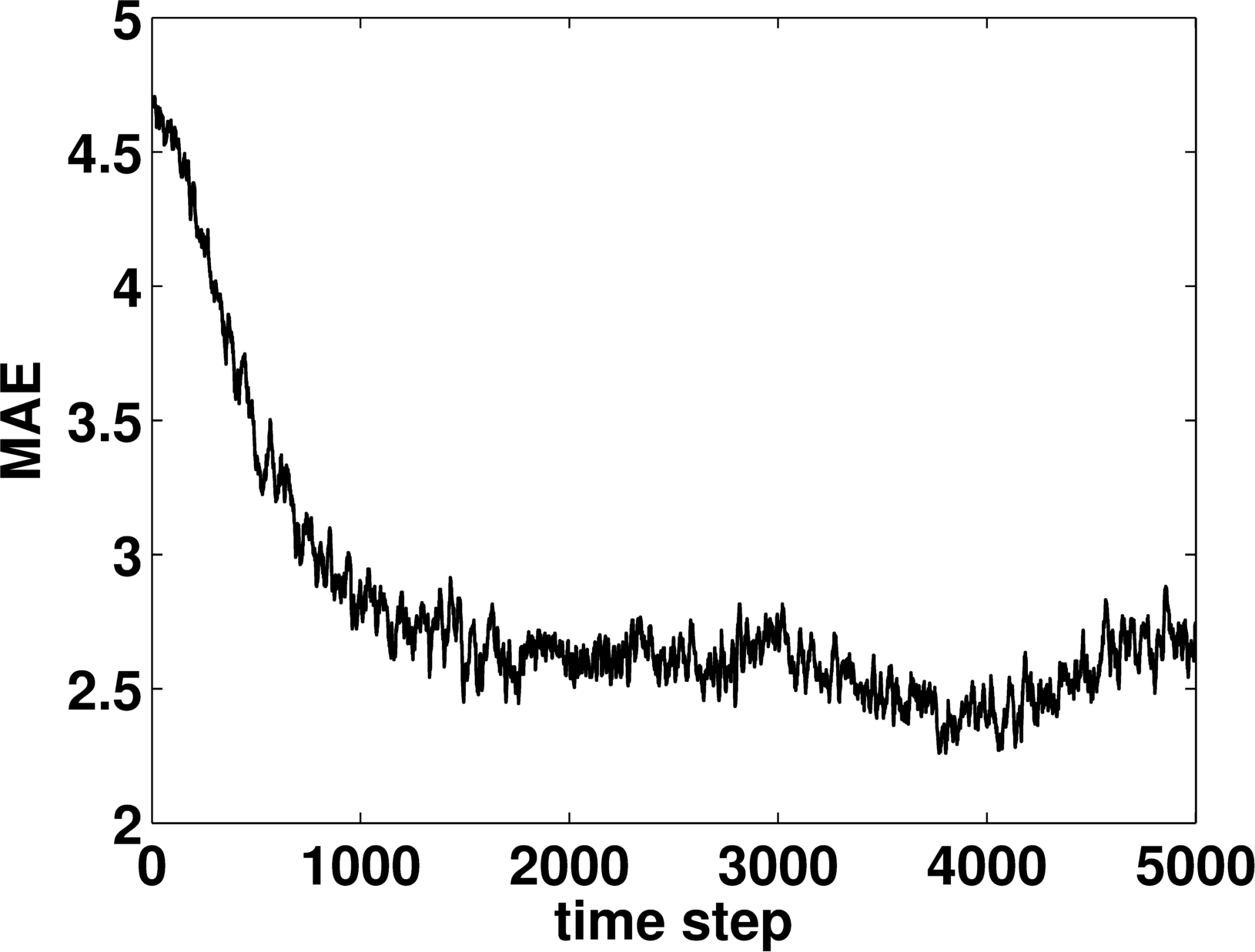}
			}
			\hfil
			\subfloat[$g=3$]
			{
				\includegraphics[width=0.3\textwidth]{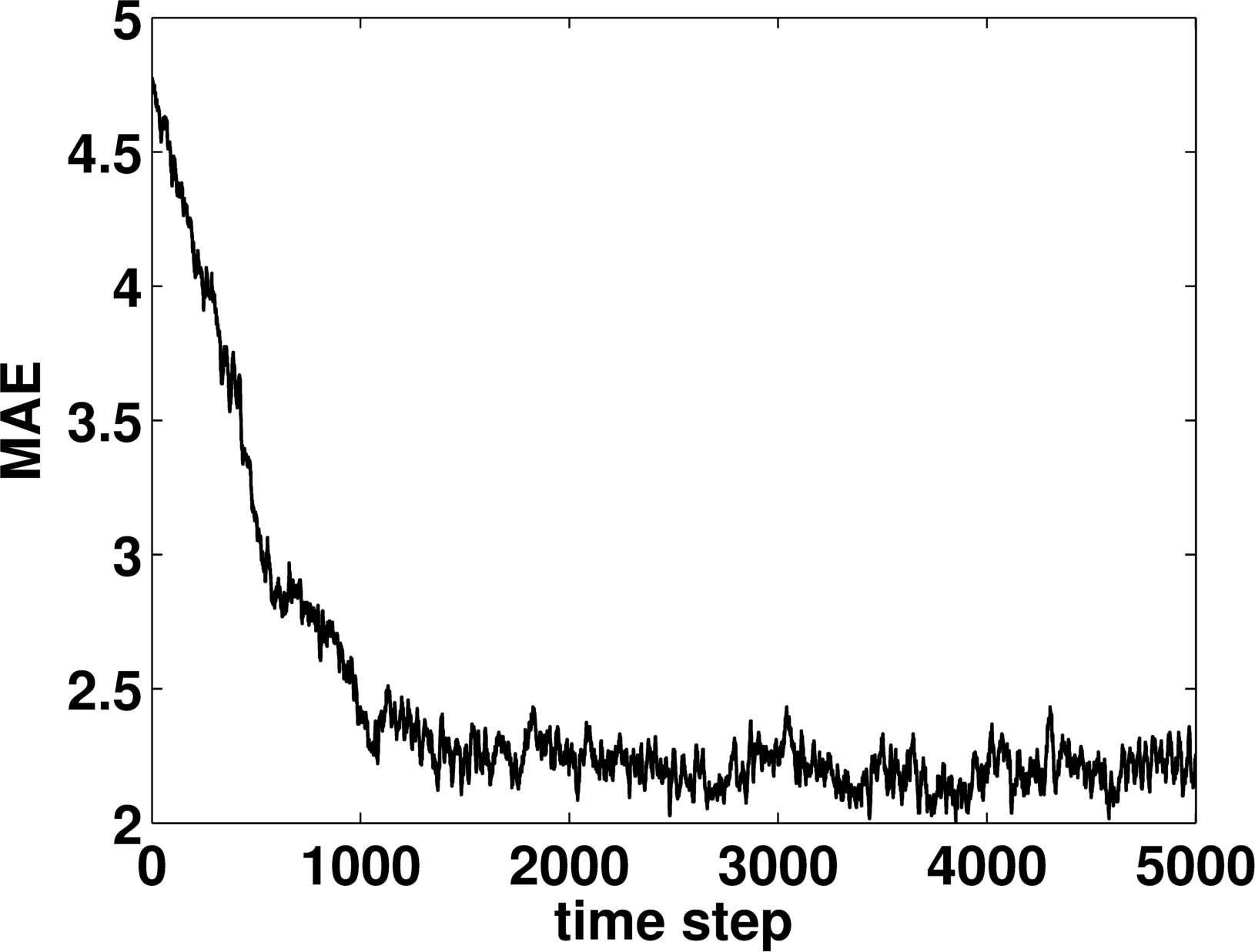}
			}
			\hfil
			\subfloat[$g=3.5$]
			{
				\includegraphics[width=0.3\textwidth]{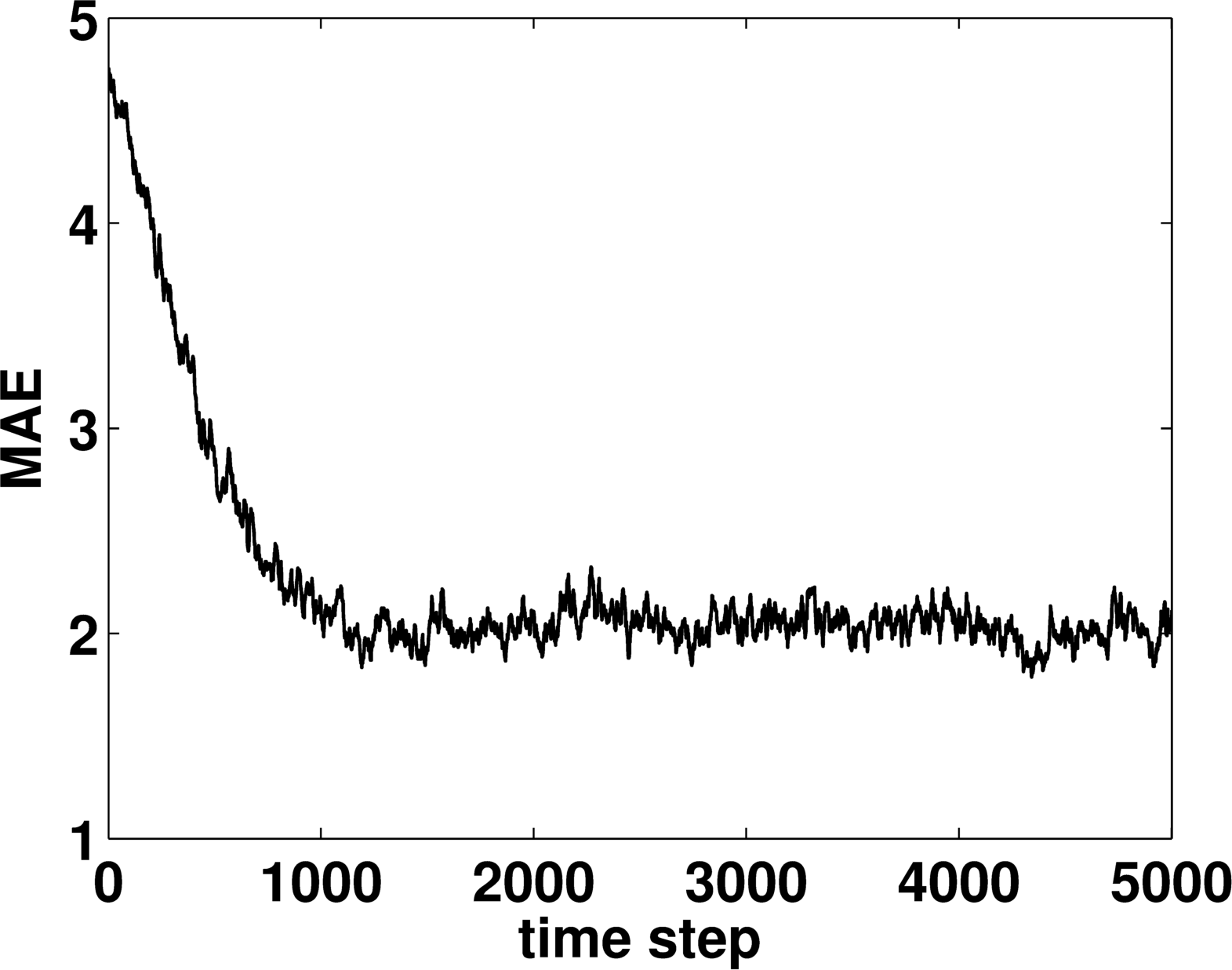}
			}
			\caption{Mean Absolute Error (MAE).}
			\label{img:mae}
		\end{figure}

		\autoref{img:mae} shows that the system has achieved equilibrium when time step is close to 2000.
		This equilibrium is indicated by the trend of Mean Absolute Error (MAE) between density profile calculated from the model \eqref{eq:rho.hidrostatik} and the simulation result.
		\autoref{img:mae} also shows that for small gravitational acceleration value ($g \le 1.5$), fluctuations still occurred although the system has achieved equilibrium.
		Especially for $g=1$ where the system is not stable enough although the trend is quite stable.
		In contrast to that, more stable fluctuations are shown by higher values of $g$.

\begin{figure}[!ht]
			\vskip -1em
			\centering
			\subfloat[$g=1$]
			{
				\includegraphics[width=0.3\textwidth]{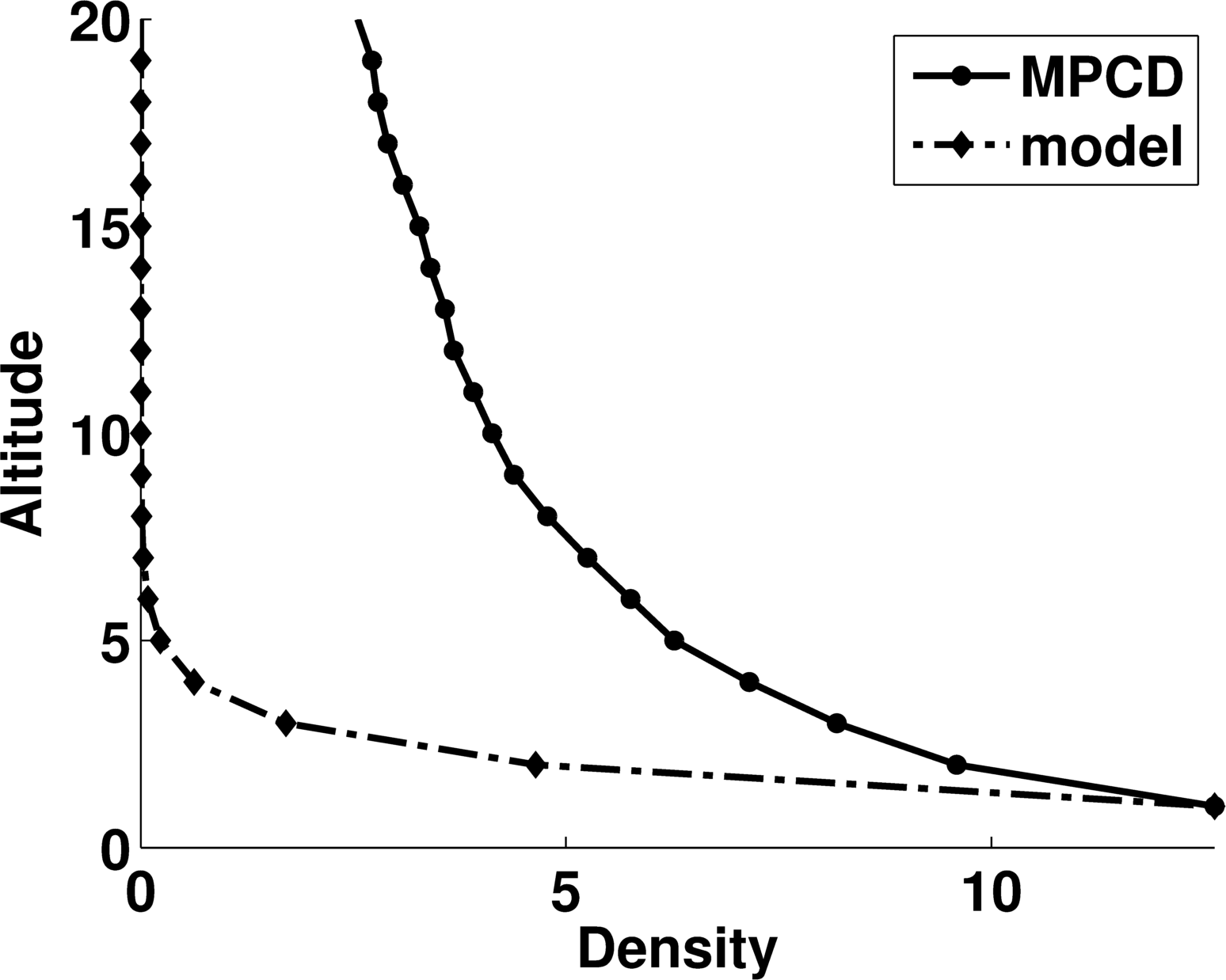}
			}
			\hfil
			\subfloat[$g=1.5$]
			{
				\includegraphics[width=0.3\textwidth]{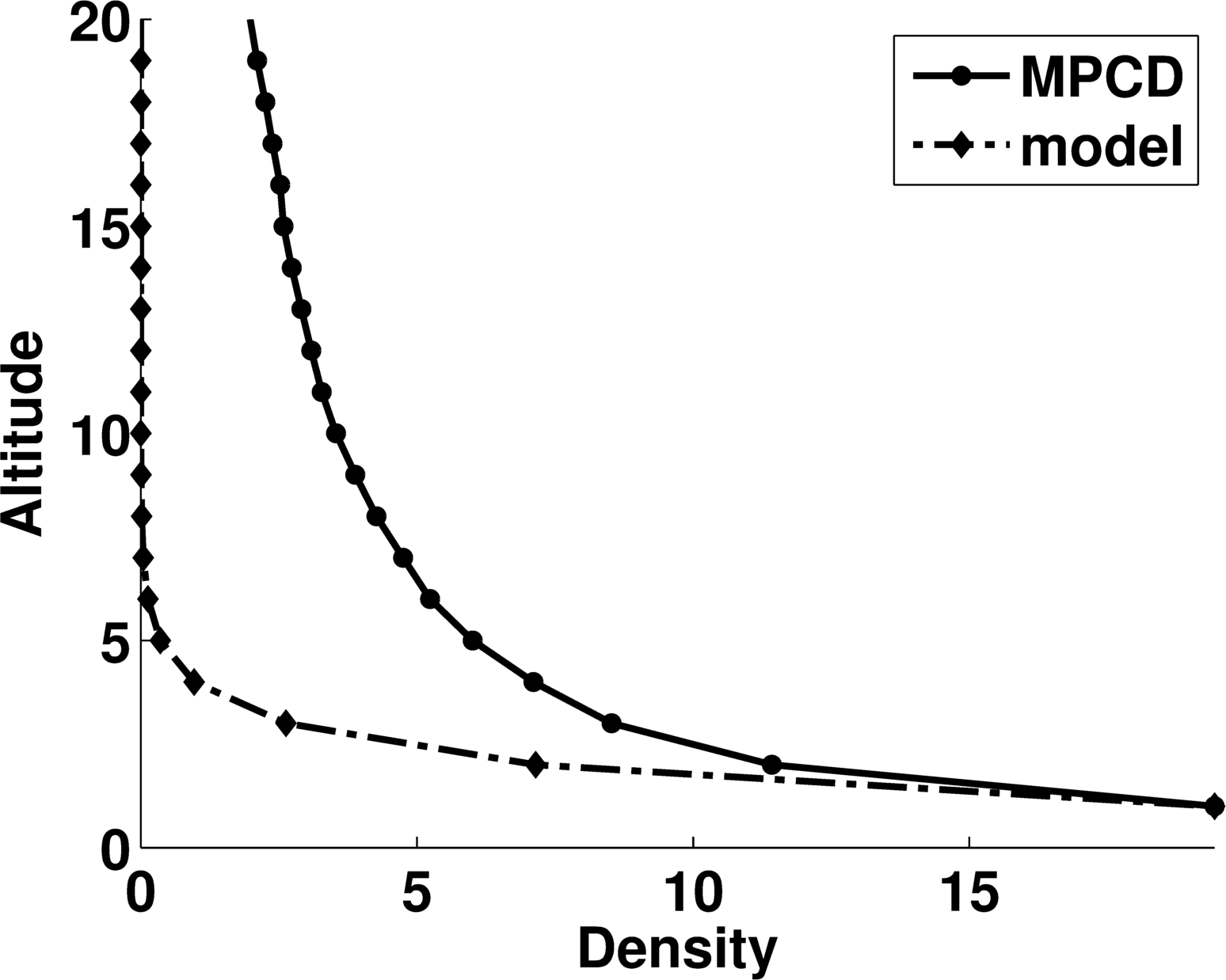}
			}
			\hfil
			\subfloat[$g=2$]
			{
				\includegraphics[width=0.3\textwidth]{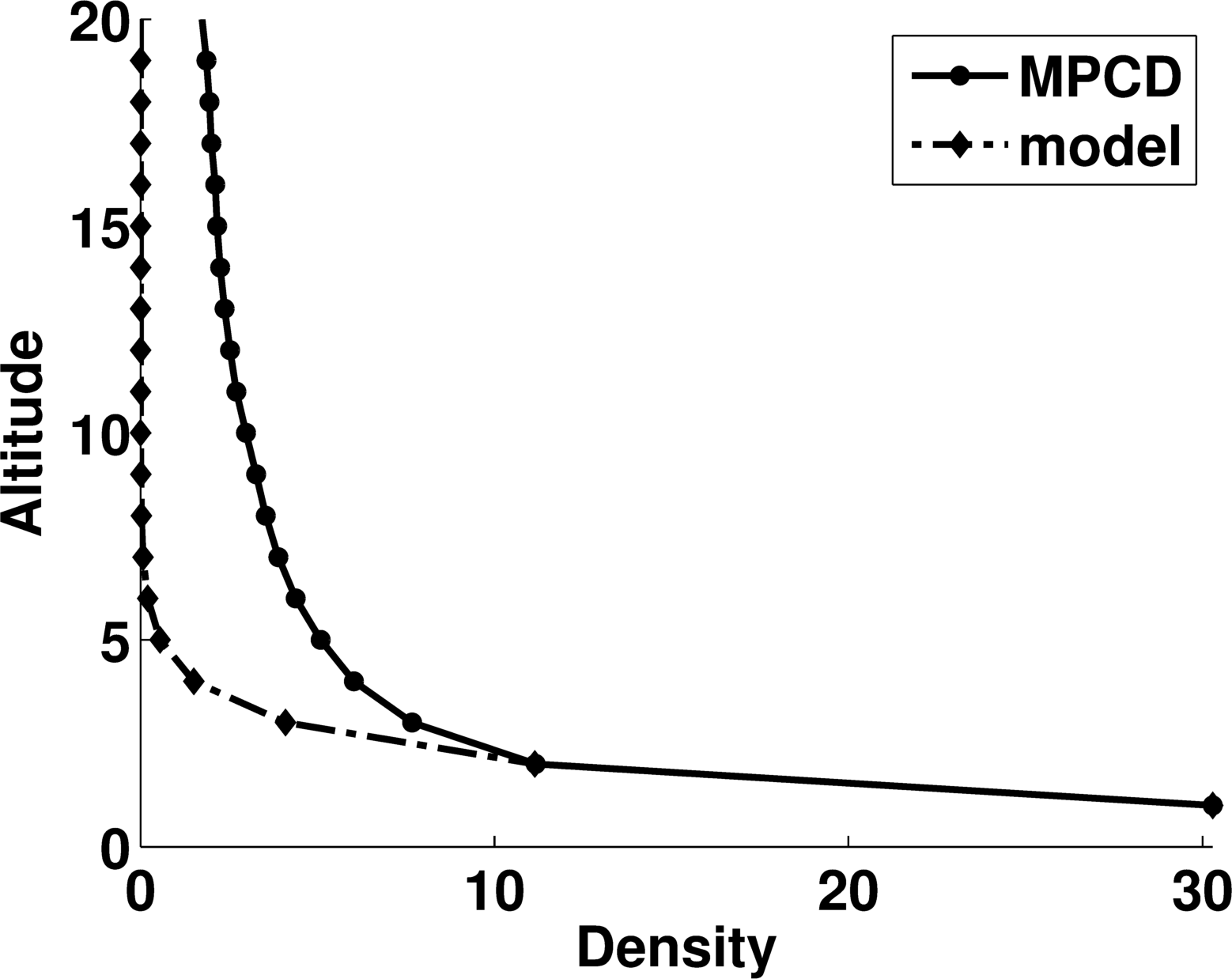}
			}
			\\
			\subfloat[$g=2.5$]
			{
				\includegraphics[width=0.3\textwidth]{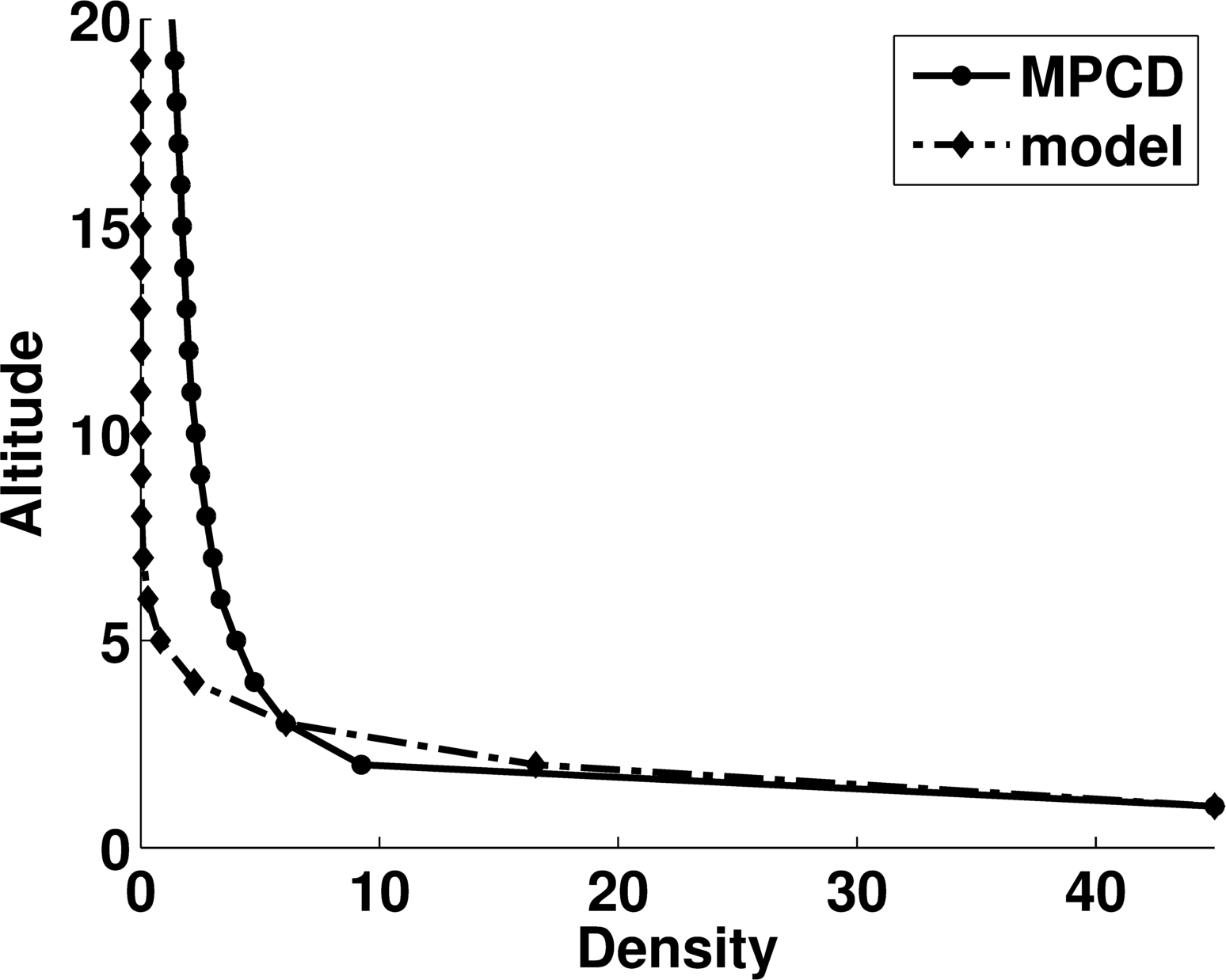}
			}
			\hfil
			\subfloat[$g=3$]
			{
				\includegraphics[width=0.3\textwidth]{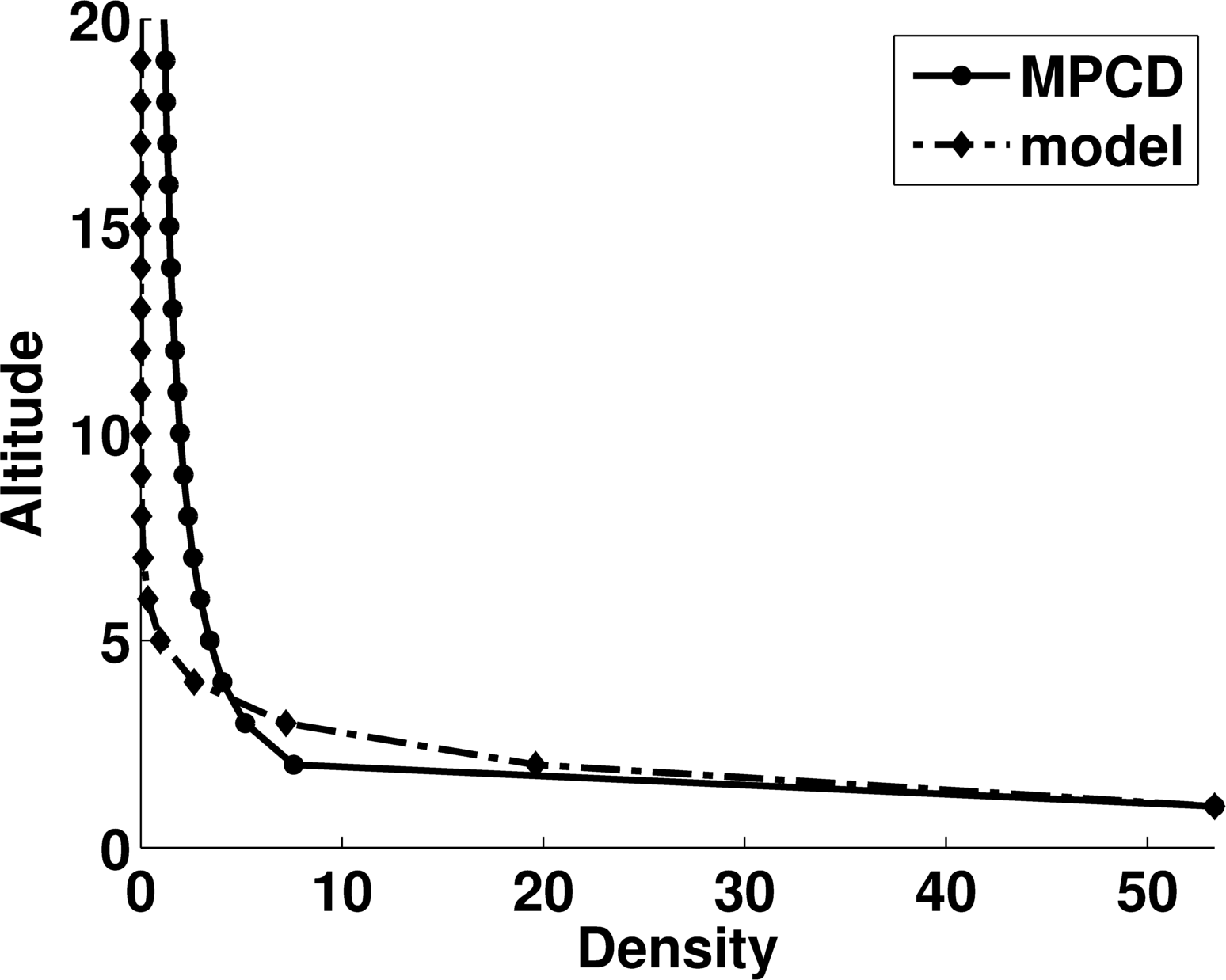}
			}
			\hfil
			\subfloat[$g=3.5$]
			{
				\includegraphics[width=0.3\textwidth]{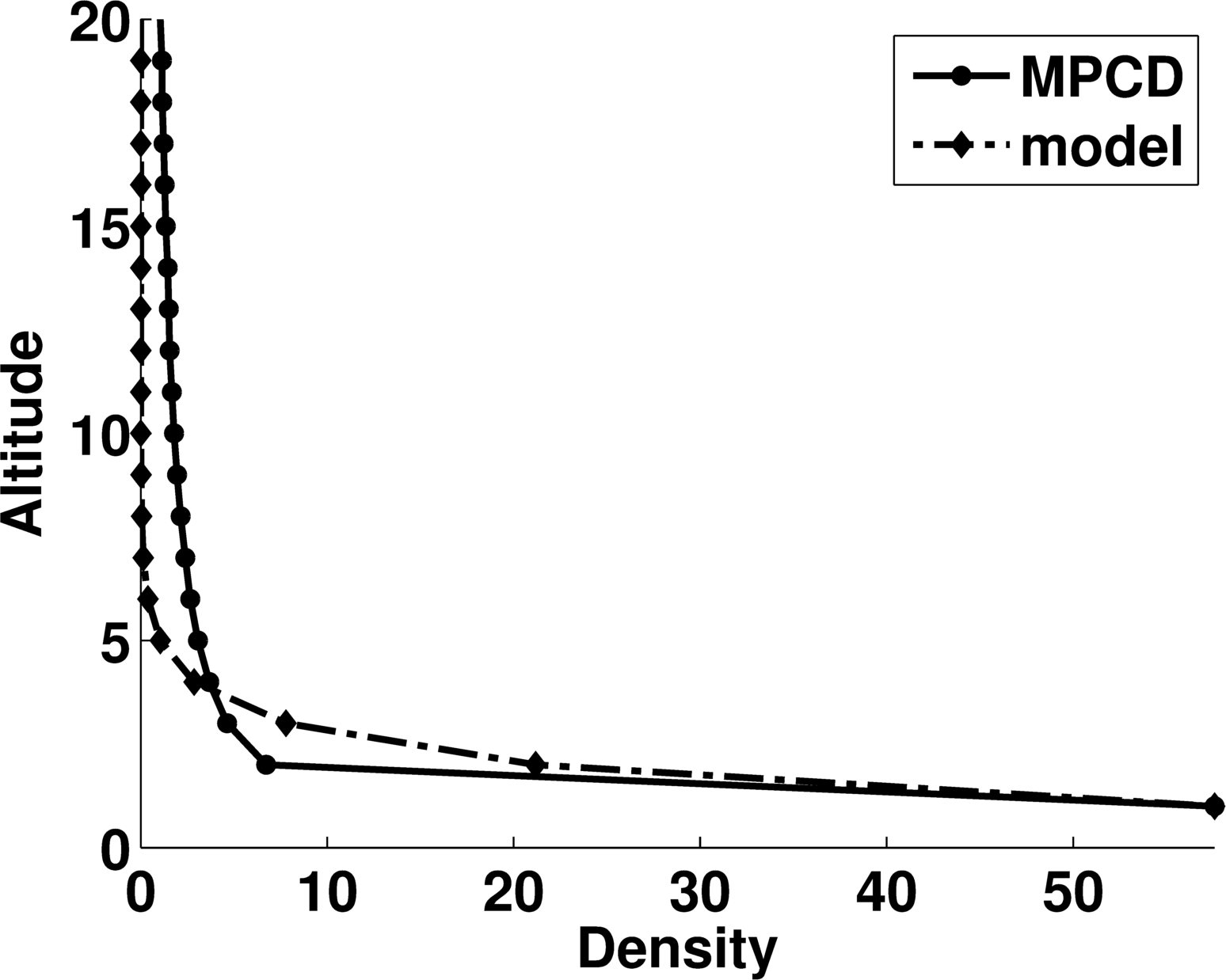}
			}
			\caption{Average density profile as a function of altitude after equilibrium: Theoretical
			prediction ($\blacklozenge$ dot-dashed line) and MPCD results ($\bullet$ solid line).}
			\label{img:density}
		\end{figure}

\begin{figure}[!ht]
			\vskip -1em
			\centering
			\subfloat[$g=1$]
			{
				\includegraphics[width=0.3\textwidth]{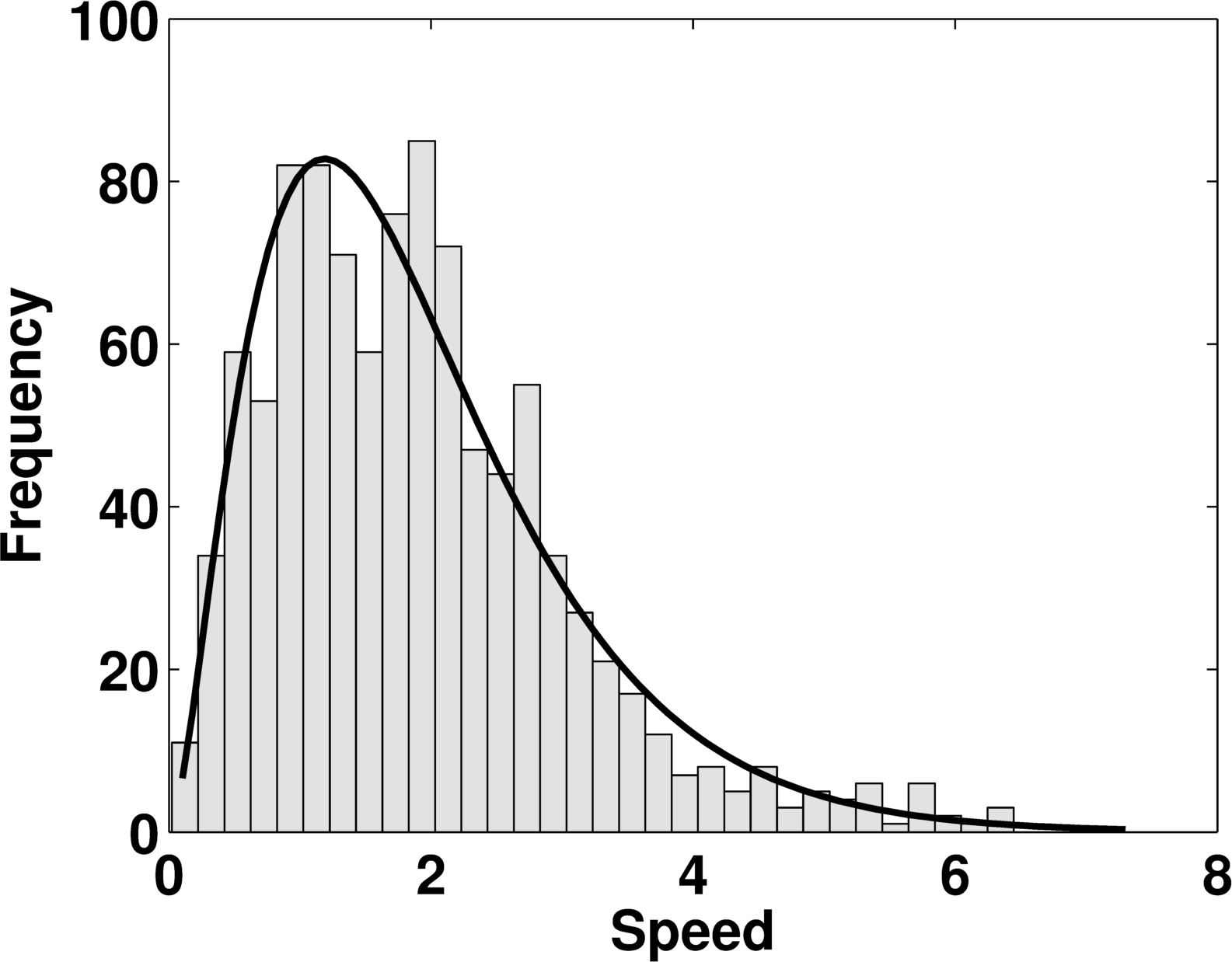}
			}
			\hfil
			\subfloat[$g=1.5$]
			{
				\includegraphics[width=0.3\textwidth]{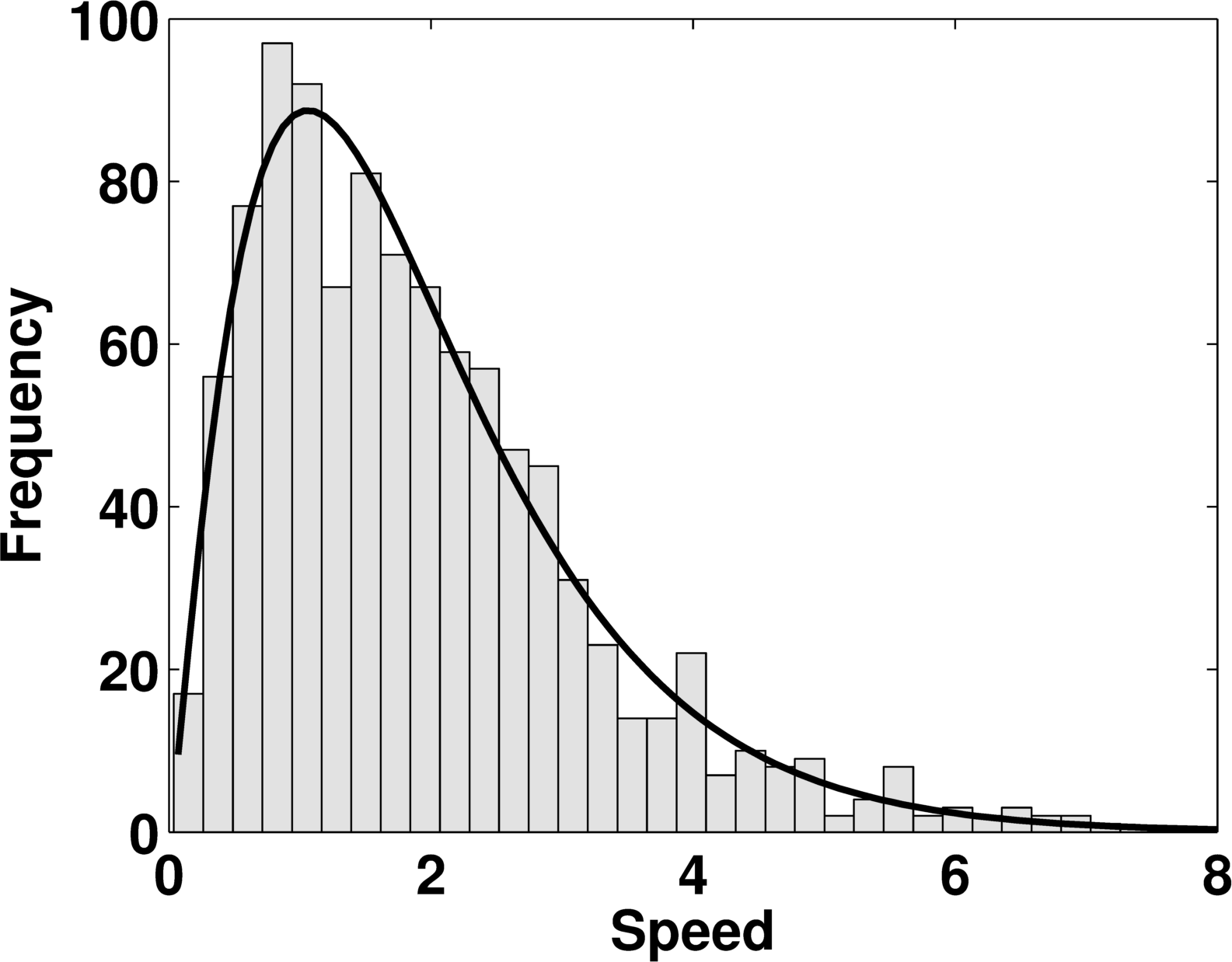}
			}
			\hfil
			\subfloat[$g=2$]
			{
				\includegraphics[width=0.3\textwidth]{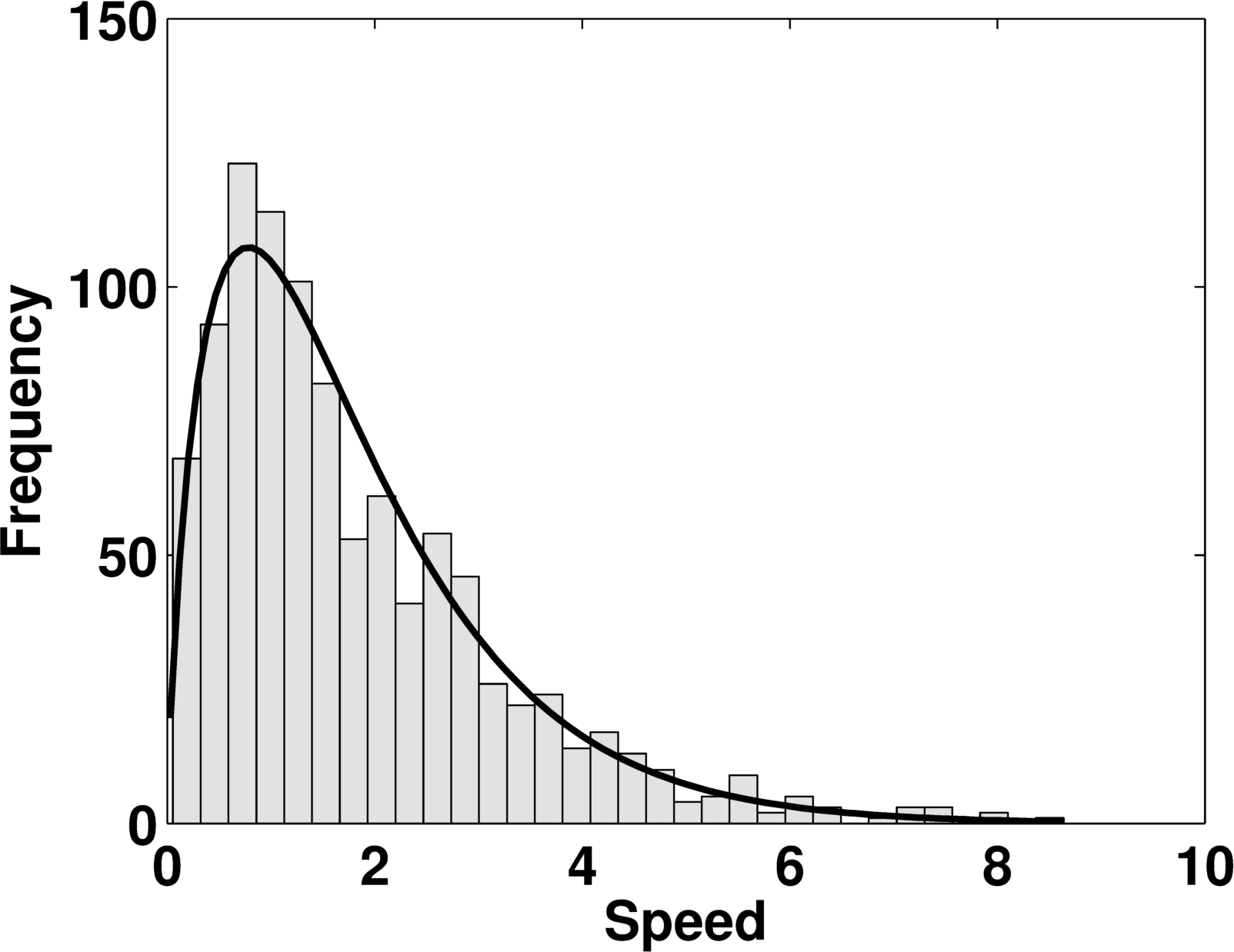}
			}
			\\
			\subfloat[$g=2.5$]
			{
				\includegraphics[width=0.3\textwidth]{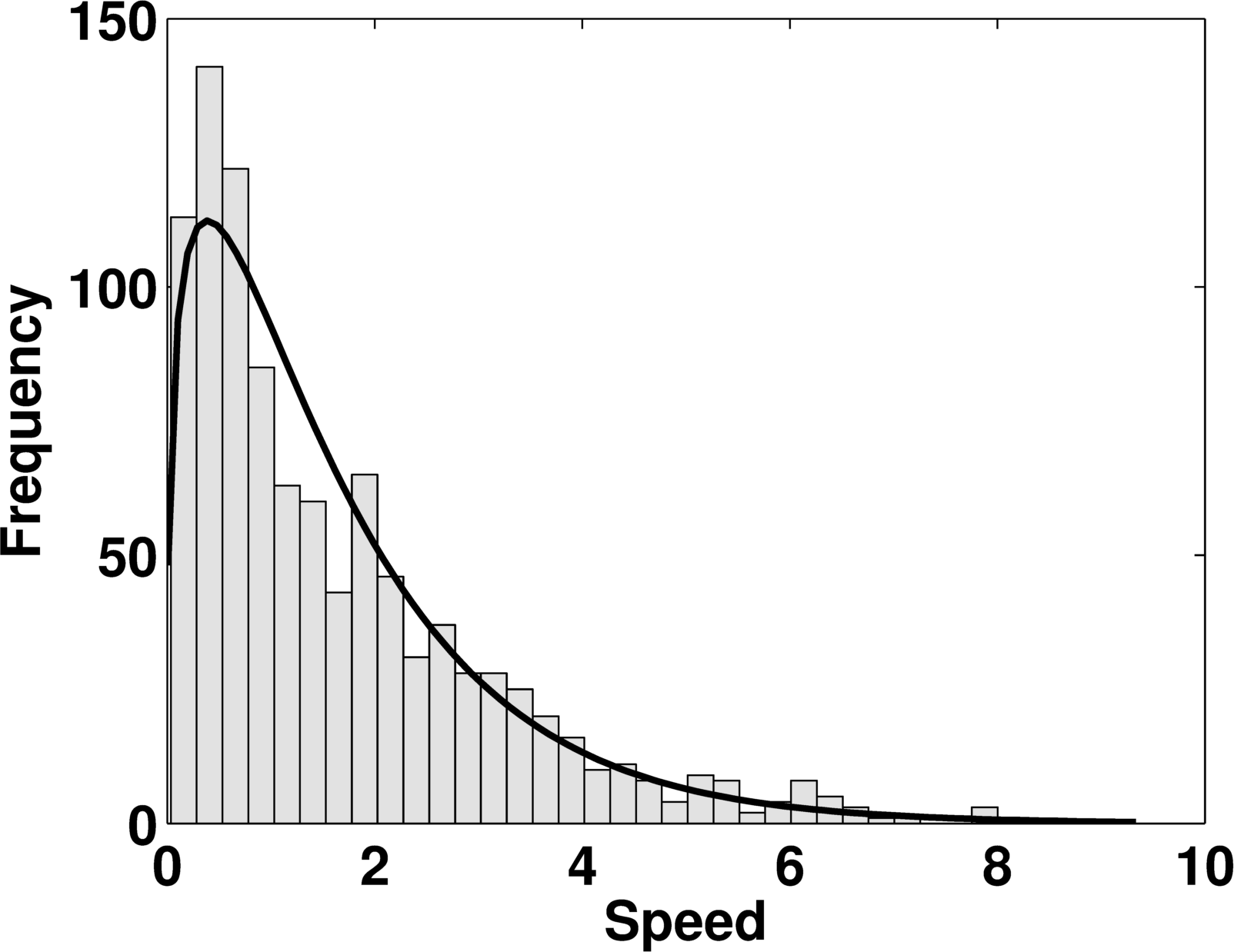}
			}
			\hfil
			\subfloat[$g=3$]
			{
				\includegraphics[width=0.3\textwidth]{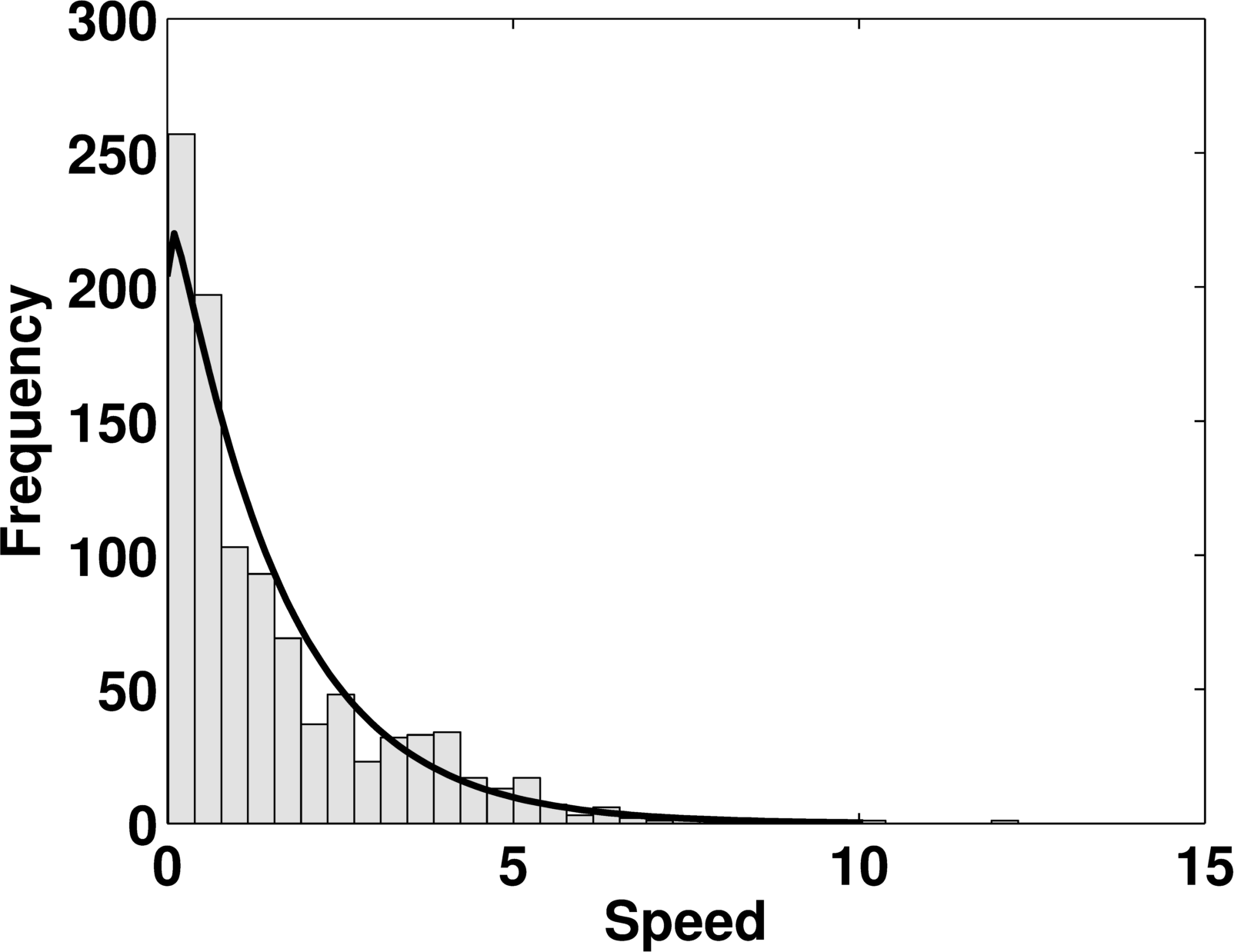}
			}
			\hfil
			\subfloat[$g=3.5$]
			{
				\includegraphics[width=0.3\textwidth]{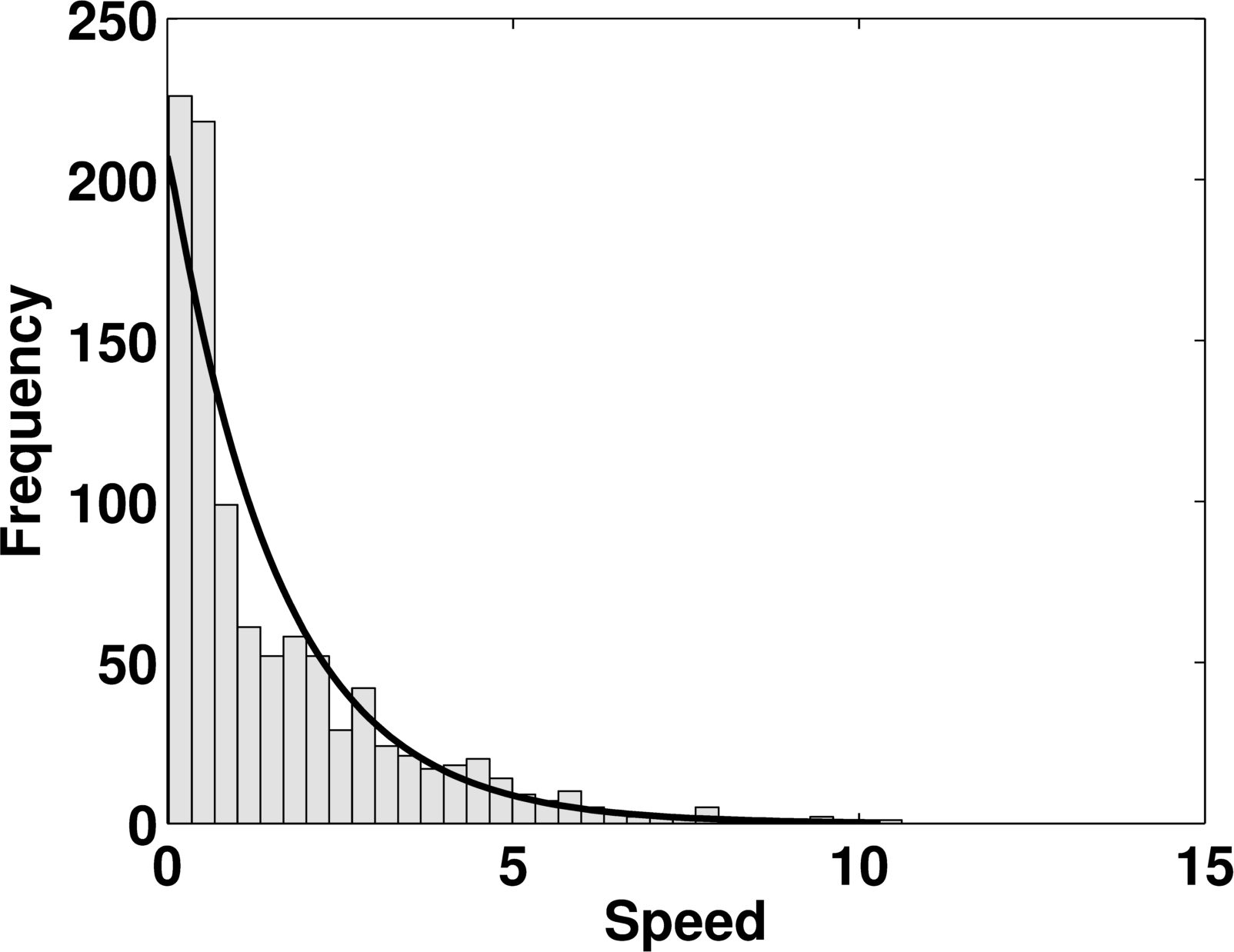}
			}
			\caption{
			Particle speed distribution
			: MPCD result (bar) and fitted (solid line).}
			\label{img:particle.speed}
		\end{figure}

		According to \autoref{img:density}, MPCD algorithm is fairly accurate in modeling hydrostatic atmospheric system.
		The dash-dotted line on the figure is calculated from equation \eqref{eq:rho.hidrostatik}, while the other one is the simulation result.
		Hydrostatic density profile can be achieved when $g \ge 2$.
		However, when $g\ge3.5$ it seems that the applied force is too big and attracts the majority of the particles to the bottom of the box, while for smaller $g$, particles are not really affected by applied force.
		These effects are also shown in the profile of particle speed distribution (\autoref{img:particle.speed}).
		Particle speeds do not obey Maxwell-Boltzmann distribution when gravitational acceleration is higher than $2.5$.

	\section{Conclusions}
		MPCD algorithm is quite fast and can be used as an alternative of lightweight (low performance computing) method for fluids simulation.
		Atmospheric hydrostatic equilibrium can be achieved by using gravitational acceleration that is proportional to the particle mass.
		Based on the results, the recommended gravitational acceleration value is about $1.5 \le g < 2.5$.
		MPCD method is capable to modeling atmospheric hydrostatic state and probably can be extended to hydrodynamic cases in atmospheric layer.

	\bibliographystyle{apalike}
	\bibliography{biblio}

\begin{thebibliography}{}

\bibitem[Allahyarov and Gompper, 2002]{allahyarov2002mesoscopic}
Allahyarov, E. and Gompper, G. (2002).
\newblock Mesoscopic solvent simulations: Multiparticle-collision dynamics of
  three-dimensional flows.
\newblock {\em Physical Review E}, 66(3):036702.

\bibitem[Bird, 1994]{bird1994molecular}
Bird, G.~A. (1994).
\newblock {\em Molecular gas dynamics and the direct simulation of gas flows}.
\newblock Clarendon Press, New York.

\bibitem[COESA, 1976]{coesa1976standard}
COESA (1976).
\newblock {\em U. S. Standard Atmosphere, 1976}.
\newblock U. S. Government Printing Office, Washington, D. C.

\bibitem[Gompper et~al., 2009]{gompper2009multi}
Gompper, G., Ihle, T., Kroll, D.~M., and Winkler, R.~G. (2009).
\newblock Multi-particle collision dynamics: A particle-based mesoscale
  simulation approach to the hydrodynamics of complex fluids.
\newblock In {\em Advanced Computer Simulation Approaches for Soft Matter
  Sciences {III}}, pages 1--87. Springer Berlin Heidelberg.

\bibitem[Hecht et~al., 2005]{hecht2005simulation}
Hecht, M., Harting, J., Ihle, T., and Herrmann, H.~J. (2005).
\newblock Simulation of claylike colloids.
\newblock {\em Physical Review E}, 72(1):011408.

\bibitem[Ihle and Kroll, 2001]{ihle2001stochastic}
Ihle, T. and Kroll, D.~M. (2001).
\newblock Stochastic rotation dynamics: a galilean-invariant mesoscopic model
  for fluid flow.
\newblock {\em Physical Review E}, 63.

\bibitem[Lee et~al., 2006]{lee2006hardware}
Lee, D.-U., Villasenor, J.~D., Luk, W., and Leong, P. H.~W. (2006).
\newblock A hardware gaussian noise generator using the box-muller method and
  its error analysis.
\newblock {\em IEEE Transactions on Computers}, 55(6):659--671.

\bibitem[Malevanets and Kapral, 1999]{malevanets1999mesoscopic}
Malevanets, A. and Kapral, R. (1999).
\newblock Mesoscopic model for solvent dynamics.
\newblock {\em Journal of Chemical Physics}, 110(17):8605--8613.

\bibitem[Malevanets and Kapral, 2000]{malevanets2000solute}
Malevanets, A. and Kapral, R. (2000).
\newblock Solute molecular dynamics in a mesoscale solvent.
\newblock {\em The Journal of Chemical Physics}, 112(16):7260--7269.

\bibitem[Nikoubashman and Likos, 2010]{nikoubashman2010flow}
Nikoubashman, A. and Likos, C.~N. (2010).
\newblock Flow-induced polymer translocation through narrow and patterned
  channels.
\newblock {\em The Journal of chemical physics}, 133(7):074901.

\bibitem[Singh et~al., 2014]{singh2014hydrodynamic}
Singh, S.~P., Huang, C.-C., Westphal, E., Gompper, G., and Winkler, R.~G.
  (2014).
\newblock Hydrodynamic correlations and diffusion coefficient of star polymers
  in solution.
\newblock {\em The Journal of chemical physics}, 141(8):084901.

\end{thebibliography}

\end{document}